\documentclass[aps,showpacs,nofootinbib,superscriptaddress,twocolumn]{revtex4}

\usepackage{mathrsfs}

\usepackage{graphicx}
\usepackage{bm}
\usepackage{amsmath}
\usepackage{amssymb}
\usepackage{hyperref}

\newcommand{\nslash}{\kern 0.2 em n\kern -0.50em /}
\newcommand{\kslash}{\kern 0.2 em k\kern -0.45em /}
\newcommand{\lslash}{\kern 0.2 em l\kern -0.50em /}
\newcommand{\pslash}{\kern 0.2 em p\kern -0.50em /}
\newcommand{\Sslash}{\kern 0.2 em S\kern -0.50em /}
\newcommand{\Pslash}{\kern 0.2 em P\kern -0.50em /}
\newcommand{\Dslash}{\kern 0.2 em D\kern -0.65em /\kern 0.15em}

\begin{document}

\title{Gluon Sivers function in a light-cone spectator model}

\author{Zhun Lu}\email{zhunlu@seu.edu.cn}
\affiliation{Department of Physics, Southeast University, Nanjing
211189, China}
\author{Bo-Qiang Ma}\email{mabq@pku.edu.cn}\affiliation{School of Physics and State Key Laboratory of Nuclear Physics and Technology, Peking University, Beijing
100871, China}
\affiliation{Collaborative Innovation Center of Quantum Matter, Beijing, China}
\affiliation{Center for High Energy Physics, Peking University, Beijing 100871, China}

\begin{abstract}
We calculate the gluon Sivers function of the proton in the valence-$x$ region using a light-cone spectator model with the presence of the gluon degree of freedom. We obtain the values of the parameters by fitting the model resulting gluon density distribution to the known parametrization. We find that our results agree with the recent phenomenological extraction of the gluon Sivers function after considering the evolution effect. We also estimate the mean transverse momentum of the gluon in a transversely polarized proton and find that it is within the range implied by the Burkardt sum rule.
\end{abstract}


\maketitle

\section{Introduction}

The Sivers function~\cite{sivers} is a leading-twist transverse momentum dependent (TMD) distribution function which describes the asymmetric distribution of unpolarized partons in a transversely polarized proton.
It is of great interest because it can give rise to azimuthal asymmetries of final-state particles in various high energy process involving a transversely polarized nucleon, also because it encodes nontrivial partonic structure in the transverse plane through spin-orbital correlation.

In recent years, the quark Sivers function $f_{1T}^{\perp \,q}(x,\bm k_\perp^2)$ has been extensively studied from both theoretical and experimental sides and much progress has been made.
Significant Sivers effect in semi-inclusive deep inelastic scattering was measured by the HERMES~\cite{Airapetian:2004tw,Airapetian:2009ae}, COMPASS~\cite{compass05,Alekseev:2010rw,Bradamante:2011xu}, and the JLab Hall A Collaborations~\cite{Qian:2011py}.
The data on the Sivers single spin asymmetries (SSAs) were further utilized by
different groups~\cite{anselmino05b,efr05,cegmms,vy05,Anselmino:2008sga}
to extract the quark Sivers functions of the proton within the TMD factorization~\cite{Ji:2004xq}.
On the other hand, there are a number of calculations on the quark Sivers function using various QCD-inspired models~\cite{Brodsky:2002plb,Yuan:2003plb,Burkardt:2003je,Lu:2004au,Cherednikov:2006zn,Lu:2006kt,
Bacchetta:2008prd,Courtoy:2008vi,Courtoy:2008dn,Pasquini:2010af}.
Furthermore, TMD evolution~\cite{collinsbook,Idilbi:2004vb,Aybat:2011ge,Aybat:2011zv} of the quark Sivers function is found to be important to consistently describe the SSA data measured at different energy scales~\cite{Aybat:2011ta}.

Compared to the quark Sivers function, the knowledge of the gluon Sivers function $f_{1T}^{\perp \,g}$ (for a review, see Ref.~\cite{Boer:2015vso}) is still limited.
Even so, model calculations of $f_{1T}^{\perp \,g}(x,\bm k_\perp^2)$ have been performed in literature, mainly focusing on the small $x$ region by means of the dipole formalism~\cite{Boer:2015pni,Boer:2016xqr}, or by employing a quark target model~\cite{Goeke:2006ef} which is different from the realistic case of proton target.
Besides, Burkardt~\cite{Burkardt:2004ur} derived a useful constraint on $f_{1T}^{\perp \,g}$, the so-called Burkardt sum rule, which states that the total transverse momentum of all partons in a transversely polarized proton should vanish.
In terms of the Sivers function, it means that the sum of the first transverse moments of all the quark, antiquark and gluon Sivers functions is zero.
Very recently, the authors of Ref.~\cite{D'Alesio:2015uta} performed a phenomenological estimate on $f_{1T}^{\perp \,g}(x,\bm k_\perp^2)$ using the midrapidity data on the transverse SSA measured in $pp \rightarrow \pi^0 X$ process at RHIC~\cite{Adare:2013ekj}.

In this work, we study the gluon Sivers function of the proton from an intuitive model concerning the gluon structure of the nucleon.
The purpose of the study is to provide information on $f_{1T}^{\perp \,g}$ in the valence-$x$ region, which is complementary to the phenomenological analysis on the experimental data as well as the dipole calculation.
The main difficulty in the calculation is how to generate the gluon degree of freedom, since in the naive parton model the proton is composed by three valence quarks.
As a first estimate, here we consider a Fock state for a transversely polarized proton that contains a gluon, and we group the three valance quark as a spectator particle.
We then present the wavefunctions for the Fock state in the light-cone formalism~\cite{Brodsky:2000ii}.
The underlying model is used to reproduce the collinear gluon density distribution $f_1^g(x)$ and to obtain the values of the parameter.
Based on this, we calculate the Sivers function using the overlap representation of the light-cone wavefunctions.
The final-state interaction necessary to produce nonzero phase for the gluon Sivers function is properly taken into account through an interaction kernal.
As a check, we also compare our numerical result with the extracted $f_{1T}^{\perp \,g}$ in Ref.~\cite{D'Alesio:2015uta} and estimate the mean transverse momentum of the gluon in a transversely polarized proton.

\section{calculation of the gluon Sivers function in an overlap representation}

The unpolarized gluon TMD distribution $f_1^g(x,\bm k_\perp^2)$ and the gluon Sivers function $f_1^{\perp g}(x,\bm k_\perp^2)$ appear in the decomposition of the correlation function $\Phi^{g}(x,\bm k_\perp;S)$~\cite{Burkardt:2004,Goeke:2006ef,Meissner:2007rx}
\begin{align} \label{eq:gc}
 &\Phi^{g}(x,\bm k_\perp;S)\nonumber\\
 &\ =\frac{1}{xP^+}\int\frac{d\xi^-}{2\pi}\,\frac{d^2\bm \xi_\perp}{(2\pi)^2}\,e^{i k\cdot \xi}\,
  \big<P;S\big|\,
  F^{+i}_a(0)\notag\\
 &\quad\ \,\times\mathcal{W}_{+\infty,ab}(0;\xi)\,
  F^{+i}_b(\xi)\,\big|P;S\big>\,
  \Big|_{\xi^+=0^+} \,
\nonumber\\
 &\ =f_1^g(x,\bm k_\perp^{\,2})
  -\frac{\epsilon^{ij}_\perp k_\perp^i S_\perp^j}{M}\,f_{1T}^{\bot g}(x,\bm k_\perp^{\,2}) \,,
\end{align}
where $F^{\mu\nu}$ is the field strength tensor of the gluon, and $\mathcal{W}_{+\infty,ab}$ is the Wilson line ensuring the gauge invariance of the correlator.
The symbol ``$+$" in the subscript denotes that the Wilson line in the operator definition of the correlator is future-pointing, which is appropriate for defining TMD distributions in SIDIS.

In Refs.~\cite{Goeke:2006ef,Meissner:2007rx}, the authors calculated the gluon TMD distributions for a quark target using perturbative QCD, in which the gluon is produced from the radiation off the parent quark.
In the case the target is a proton, the presence of the gluon degree of freedom is not obvious.
The minimum Fock state for the proton that containing gluon is $|qqqg\rangle$.
As the four-body system is very complicated, here we resort to a more phenomenological approach to assume that the three quarks can be grouped into a spectator particle.
Thus, in this model in which the degree of freedom of a gluon is present, the proton can be viewed as a composite system formed by a gluon and a spectator particle $X$:
\begin{align}
|P; S\rangle \mapsto |g^{s_{\rm{g}}} \,X^{s_{\rm{X}}}(uud)\rangle\,,
\end{align}
with $s_{\rm{g}}$ and $s_{\rm{X}}$ the spin indices for the gluon and the spectator particle.
In principle the spectator has the spin quantum number $s_{\rm{X}}=1/2$ or $3/2$.
In this work we only consider the spin-1/2 component, that is, we assume that the contribution from the spin-$3/2$ component is negligible for simplicity.
Therefore, in the case the gluon is the active parton, the Fock-state expansion of proton with $J_z=+1/2$ has the following possible form:
\begin{align}
&\hspace{-0.5cm}\left|\Psi^{\uparrow}_{\rm two \ particle}(P^+, \bm P_\perp = \bm
0_\perp)\right>
\ =\
\int\frac{{\mathrm d}^2 {\bm k}_{\perp} {\mathrm d} x }{16
\pi^3{\sqrt{x(1-x)}}}
\displaybreak[0]\nonumber\\
&\times
\Big[ \
\psi^{\uparrow}_{ +1 \,+\frac{1}{2}}(x,{\bm k}_{\perp})\,
\left|+1\,, +\frac{1}{2} \, ;\,\, xP^+\, ,\,\, {\bm k}_{\perp}\right>\nonumber \\
&+ \psi^{\uparrow}_{+1\,-\frac{1}{2}} (x,{\bm k}_{\perp})\,
\left| +1\,, -\frac{1}{2}\, ;\,\, xP^+\, ,\,\, {\bm k}_{\perp}\right>
\displaybreak[0]\nonumber\\
&+\psi^{\uparrow}_{ -1\, +\frac{1}{2}}(x,{\bm k}_{\perp})\,
\left| -1\, ,+\frac{1}{2}\, ;\,\, xP^+\, ,\,\, {\bm k}_{\perp}\right> \displaybreak[0]\nonumber\\
&+\psi^{\uparrow}_{-1\,-\frac{1}{2}} (x,{\bm k}_{\perp})\,
\left| -1\,,-\frac{1}{2}\, ;\,\, xP^+\, ,\,\, {\bm k}_{\perp}\right>\ \Big]
\ , \label{eq:fock1}
\end{align}
where $\psi^{\uparrow}_{ s_{\rm g}^z\,s_{\rm X}^z}(x,{\bm k}_{\perp})$ are the wavefunctions corresponding to the two-particle states $|s_{\rm g}^z, s_{\rm X}^z; \ xP^+, {\bm
k}_{\perp} \rangle$.
Here $s_{\rm g}^z$ and $s_{\rm X}^z$ denote the
$z$-components of the spins of the constituent gluon and spectator,
respectively, and $x$ is the longitudinal momentum fraction of the gluon.
Motivated by the wavefunction of the electron Fock state~\cite{Brodsky:2000ii}, the Fock state of which is composed of a spin-1 photon and a spin-$1/2$ electron, we propose that the light-cone wavefunctions appearing in Eq.~(\ref{eq:fock1}) have the following forms
\begin{equation}
\left
\{ \begin{array}{l}
\psi^{\uparrow}_{+1 \,+\frac{1}{2}} (x,{\bm k}_{\perp})=-{\sqrt{2}}
\frac{(-k_{\perp}^1+{\mathrm i} k_{\perp}^2)}{x(1-x)}\,
\varphi \ ,\\
\psi^{\uparrow}_{+1 \,-\frac{1}{2}} (x,{\bm k}_{\perp})=-{\sqrt{2}}
\left(M-{M_X\over (1-x)}\right)\,
\varphi \ ,\\
\psi^{\uparrow}_{-1\,+\frac{1}{2}} (x,{\bm k}_{\perp})=-{\sqrt{2}}
\frac{(+k_{\perp}^1+{\mathrm i} k_{\perp}^2)}{x }\,
\varphi \ ,\\
\psi^{\uparrow}_{-1\,-\frac{1}{2}} (x,{\bm k}_{\perp})=0\ ,
\end{array}
\right.
\label{eq:wf1}
\end{equation}
where $M$, and $M_X$ are the masses of the proton and the spectator state, respectively, and
$\varphi\equiv\varphi (x,{\bm k}_{\perp})$ is the wavefunction in the momentum space
\begin{equation}
\varphi (x,{\bm k}_{\perp})=\frac{ \lambda/\sqrt{x}}{M^2-({\bm
k}_{\perp}^2+M_g^2)/x-({\bm k}_{\perp}^2+M_X^2)/(1-x)}\,,
\label{wfdenom}
\end{equation}
with $\lambda$ the coupling of the nucleon-gluon-spectator vertex, and $M_g$ the gluon mass.
In principle gluon is a massless gauge boson.
Here we keep $M_g$ in our formula following the convention used in Ref.~\cite{Brodsky:2000ii}.
As shown in the next section, we fix $M_g=0$ GeV in our numerical calculation.

Although the wavefunctions in Eq.~(\ref{eq:wf1}) are similar to those of the electron, there are several differences between them.
The first one is that the mass of the spectator particle $M_X$ could be different from the mass of the proton, while in the electron case, the spectator fermion is the same as the electron.
The second one is that coupling for the electron wavefunction $e$ is a constant, whereas the coupling $\lambda$ is not necessary to be a constant, since there is nonperturbative color interaction involved in the nucleon-gluon-spectator vertex.
In order to simulate the nonperturbative physics for the vertex, we adopt the Brodsky-Hwang-Lepage prescription~\cite{Brodsky:1980vj} for the coupling $\lambda$:
\begin{equation}
\lambda \rightarrow N_\lambda\exp(-{\mathcal{M}^2\over 2\beta_1^2})\,.\label{eq:nlam}
\end{equation}
Here $N_\lambda$ is a constant parameter which represents the strength of the proton-gluon-spectator vertex, $\beta_1$ a cutting off parameter, and $\mathcal{M}$
the invariant mass of the two particle system:
\begin{equation}
\mathcal{M}^2= {\bm k_\perp^2 + M_g^2\over x}  +{\bm k_\perp^2 + M_X^2\over 1-x} .
\label{eq:m2}
\end{equation}

Similarly, the Fock-state expansion for a proton with $J_z=-1/2$ has the form
\begin{align}
&\hspace{-0.5cm}\left|\Psi^{\downarrow}_{\rm two \ particle}(P^+, \vec P_\perp =
\vec 0_\perp)\right>
\ =\
\int\frac{{\mathrm d}^2 {\bm k}_{\perp} {\mathrm d} x }{16
\pi^3{\sqrt{x(1-x)}}}
\displaybreak[0]\nonumber\\
&\times
\Big[\
\psi^{\downarrow}_{+1\,+\frac{1}{2}}(x,{\bm k}_{\perp})\,
\left|+1\,, +\frac{1}{2} \, ;\,\, xP^+\, ,\,\, {\bm k}_{\perp}\right>
\displaybreak[0]\nonumber\\
& + \psi^{\downarrow}_{+1\,-\frac{1}{2}}(x,{\bm k}_{\perp})\,
\left| +1\,,-\frac{1}{2}\, ;\,\, xP^+\, ,\,\, {\bm k}_{\perp}\right>
\displaybreak[0]\nonumber\\
&+  \psi^{\downarrow}_{-1\,+\frac{1}{2}\,}(x,{\bm k}_{\perp})\,
\left| -1\, ,+\frac{1}{2}\, ;\,\, xP^+\, ,\,\, {\bm k}_{\perp}\right>
\nonumber\\
&+\psi^{\downarrow}_{-1\,-\frac{1}{2}}(x,{\bm k}_{\perp})\,
\left|-1\,, -\frac{1}{2}\, ;\,\, xP^+\, ,\,\, {\bm k}_{\perp}\right>\ \Big]
\ ,
\label{eq:fock2}
\end{align}
where
\begin{equation}
\left
\{ \begin{array}{l}
\psi^{\downarrow}_{+1\,+\frac{1}{2}} (x,{\bm k}_{\perp})=0\ ,\\
\psi^{\downarrow}_{+1\,-\frac{1}{2}} (x,{\bm k}_{\perp})=-{\sqrt{2}}
\frac{(-k_{\perp}^1+{\mathrm i} k_{\perp}^2)}{x }\,
\varphi \ ,\\
\psi^{\downarrow}_{-1\,+\frac{1}{2}} (x,{\bm k}_{\perp})=-{\sqrt{2}}
\left(M-{M_X\over (1-x)}\right)\,
\varphi \ ,\\
\psi^{\downarrow}_{-1\,-\frac{1}{2}} (x,{\bm k}_{\perp})=-{\sqrt{2}}
\frac{(+k_{\perp}^1+{\mathrm i} k_{\perp}^2)}{x(1-x)}\,
\varphi \ .
\end{array}
\right.
\label{eq:wf2}
\end{equation}

Having the light-cone wavefunctions of the proton, we can directly calculate the unpolarized gluon TMD distribution $f_1^g(x,\bm k_\perp^2)$ using the overlap representation
\begin{align}
f_1^{g}(x,\bm k_\perp^2)&=\sum_{s_{\rm g}^z\,s_{\rm X}^z}\int  {\textrm{d}^2 \bm k_\perp\over 16\pi^3} \,\psi^{\uparrow\star}_{s_{\rm g}^z\,s_{\rm X}^z}(x,\bm k_\perp) \, \psi_{s_{\rm g}^z\,s_{\rm X}^z}^\uparrow (x,\bm k_\perp^\prime),\nonumber
 \end{align}
which yields the following result
\begin{align}
f_1^g(x,\bm k_\perp^2) &= {2N^2_\lambda \over 16\pi^3 x} \exp\left(-{\bm k_\perp^2+L_2^2(x)\over \beta_1^2\,x\,(1-x)}\right)\nonumber \\
&\times {[(1+(1-x)^2)\bm k_\perp^2 + x^2((1-x)M-M_X)^2] \over (\bm k_\perp^2+L^2_1(x))^2},
 \end{align}
with
\begin{align}
L^2_1(x) &= (1-x)M_g^2+x M_X^2 -x(1-x)M^2,\nonumber \\
L^2_2(x) &= (1-x)M_g^2+x M_X^2. \nonumber
\end{align}

After the transverse momentum $\bm k_\perp$ is integrated out, the unpolarized distribution of the gluon has the form
\begin{align}
f_1^g(x) &= {N_\lambda^2  \over 8\pi^2 x}
\exp\left(-{2aL_2^2(x)}\right) \nonumber\\
& \times
\left[ {x^2((1-x)M-M_S)^2-(1+(1-x)^2)L_1^2(x)\over  L_1^2(x)}\right.\nonumber\\
&+\bigg((1+(1-x)^2)\left(2aL_1^2(x)+1\right)\nonumber\\
&  -\,  x^2((1-x)M-M_S)^2 \bigg)\nonumber\\
&\left.\times\exp\left({2aL_1^2(x)}\right)\Gamma\left(0,{2aL_1^2(x)}\right)\right],
\end{align}
where $a=1/\left(2x(1-x)\beta_1^2\right)$, and
\begin{align}
\Gamma(n,x) = \int_x^\infty dt\, {e^{-t} \over t^{1-n} }
\end{align}
is the incomplete gamma function.

\begin{figure*}
\centering
\scalebox{0.35}{\includegraphics*[70pt,30pt][725pt,520pt]{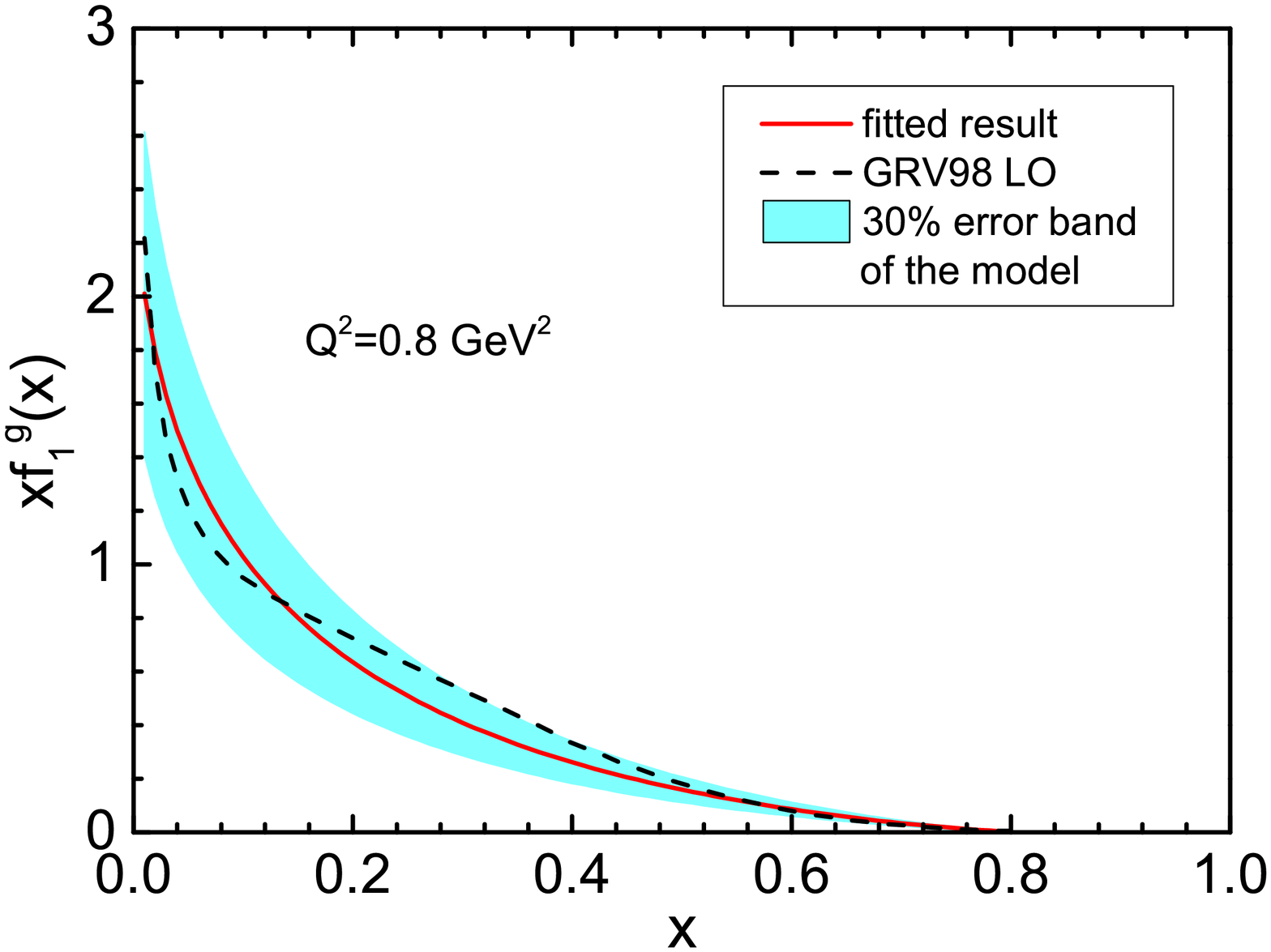}}~~
\scalebox{0.35}{\includegraphics*[70pt,30pt][725pt,520pt]{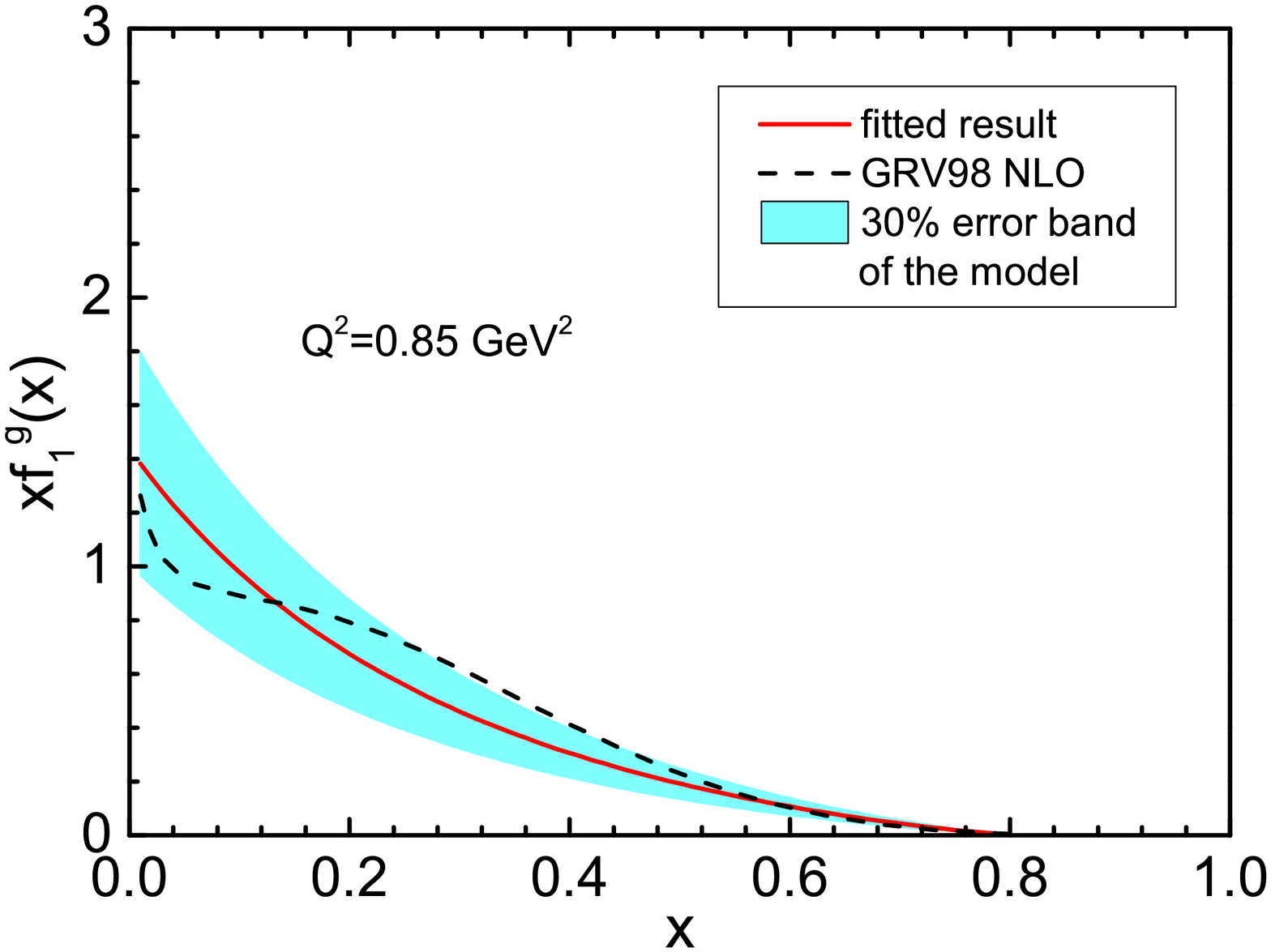}}
\caption{The fitting of the model results to the GRV98 LO (left panel) and NLO (right panel) gluon density distribution $f_1^g(x)$. The solid and dashed lines represent the model results and the GRV98 parametrizations, respectively.
The band corresponds to the 30\% error assigned to the
model.}
\label{fig:f1g}
\end{figure*}

In the overlap representation, the gluon Sivers function may be calculated from the expression~\cite{Lu:2006kt,Bacchetta:2008prd}
\begin{align}
&{ k_\perp^1 - i\,k_\perp^2\over 2M}f_1^{\perp\, g}(x,\bm k_\perp)=i\sum_{s_{\rm g}^z\,s_{\rm X}^z}\int  {d^2 \bm k_\perp^\prime\over 16\pi^3} \,\psi^{\uparrow\star}_{s_{\rm g}^z\,s_{\rm X}^z}(x,\bm k_\perp)
\nonumber\\
& \times G(x,\bm k_\perp, \bm k_\perp^\prime) \,\psi_{s_{\rm g}^z\,s_{\rm X}^z}^\downarrow (x,\bm k_\perp^\prime),\label{eq:repsiv}
\end{align}
where $ G(x,\bm k_\perp, \bm k_\perp^\prime)$ is the interaction kernel which simulates the gluon rescattering between the active parton and the spectator.
Originally the overlap representation is applied to calculate various form factors of the nucleon, as well as the nucleon anomalous magnetic moment.
Recently it has also been adopted to calculate the quark Sivers functions~\cite{Lu:2006kt,Bacchetta:2008prd} and quark Boer-Mulders function~\cite{Bacchetta:2008prd} in the spectator model.
In this work we adopt the form of $G(x,\bm k_\perp, \bm k_\perp^\prime)$ as follows:
\begin{align}
G(x,\bm k_\perp, \bm k_\perp^\prime) =  {-iC_A \alpha_S(\bm k_\perp^L-\bm k_\perp^{\prime \,L})\over 4\pi x (\bm k_\perp - \bm k_\perp^\prime)^2}\,, \label{eq:kernel}
\end{align}
which is extracted from the calculation of the gluon sivers function in the quark target model~\cite{Goeke:2006ef,Meissner:2007rx}.
Of course the final state interaction kernel should be model dependent.
Here we assume that the kernel in our spectator model is the same as that in the quark target model, since in the quark target model the spectator is also a spin-1/2 particle.

Substituting Eq.~(\ref{eq:kernel}) into Eq.~(\ref{eq:repsiv}) and performing the integration over $\bm k_\perp^\prime$, we arrive at the result of the gluon Sivers function in the spectator model:
\begin{align}
f_{1T}^{\perp\, g}(x,k_T^2)& = {(1-x) C_A \alpha_S N_\lambda^2  \over 8\pi^3}  {M\left((1-x)M-M_S\right)\over
\bm k_\perp^2(\bm k_\perp^2+L_1^2(x))} \nonumber\\
&  \times \left(\Gamma(0,aL_1^2(x))-\Gamma(0,a(\bm k_\perp^2+L_1^2(x)))\right)\nonumber\\
&\times \exp\left(-a\left(2L_2^2(x)+\bm k_\perp^2-L_1^2(x)\right) \right) \,,
\end{align}
where we have used the following integration formula:
\begin{align}
&\int d^2 \bm k_\perp^\prime \exp(-a\bm k_\perp^{\prime \,2})
{\bm k_\perp^2 - \bm k_\perp \cdot \bm k_\perp^\prime \over (\bm k_\perp^\prime - \bm k_\perp)^2(\bm k_\perp^{\prime\,2} +b)^m} \nonumber\\
&= \pi \exp(ab) \left(\Gamma(1-m,ab)-\Gamma\left(1-m,a\right(\bm k_\perp^2 +b\left)\right)\right)\,.
\end{align}

\begin{table}[t]
\centering
\begin{tabular}{c|c|c}
  \hline
  ~Parameters~ & ~Fit 1 (LO)~ &~  Fit 2 (NLO)   \\
\hline\hline
 $N_\lambda$ & ~$5.026 ~$ & ~$5.865~$ \\
  \hline
  $M_X$ (GeV)
  & $0.943 $ & $1.023$  \\
  \hline
  $\beta_1$ (GeV) &
  $2.092 $  &  $2.307 $ \\
  \hline
   $M_g$ (GeV) & ~0 (fixed) ~ & ~0 (fixed) ~  \\
  \hline
  $Q_0^2$ (GeV$^2$) & ~0.80  ~ & ~0.85~ \\
  \hline
\end{tabular}
\caption{Values of the parameters in the spectator model obtained from fitting the model to the LO (second column) and NLO (third column) sets of the GRV98 gluon PDF.}\label{table1}
\end{table}

\section{Numerical results}

To present the numerical result for the gluon Sivers function, we need to specify the values of the parameters $N_\lambda$, $M_X$, $\beta_1$ and $M_g$ in our model.
As the unpolarized gluon distribution $f_1^g(x)$ in the valence region is fairly known, we fit the spectator model result of $f_1^g(x)$ to the existed parametrization for $f_1^g(x)$ to determine the values of the above parameters.
We will choose the GRV98 parametrization~\cite{Gluck:1998xa} to perform two different fits for comparison.
We note that the same parametrization has been applied in Ref.~\cite{D'Alesio:2015uta} to extract the gluon Sivers function.
We select 80 data points in the interval $0.01 \le x \le  0.80$ with a step value of $0.01$.
We have tried to include the gluon PDF in the smaller $x$ region or in the larger $x$ region in the fit,
However, we find that in this case a good fit cannot be achieved.
This indicates that our model is applicable in the $x$ region which is not so small and not so large.
In the first fit (denoted as fit 1) we apply the leading order (LO) set of GRV98 gluon PDF, while in the second fit (denoted as fit 2) we adopt the next-to-leading order (NLO) set of GRV98 gluon PDF.

\begin{figure*}
\centering
\scalebox{0.35}{\includegraphics*[65pt,30pt][720pt,515pt]{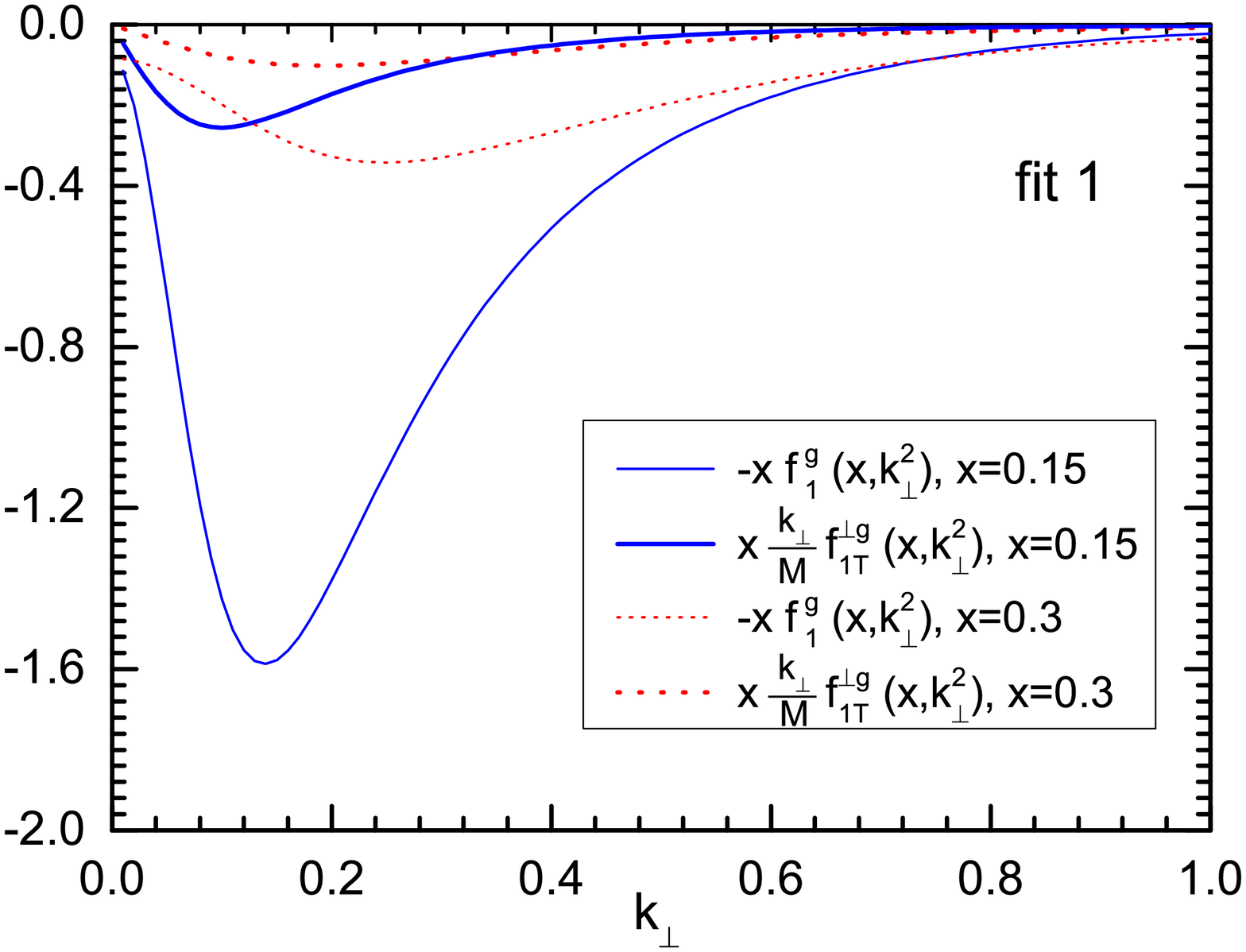}}~~
\scalebox{0.35}{\includegraphics*[65pt,30pt][720pt,515pt]{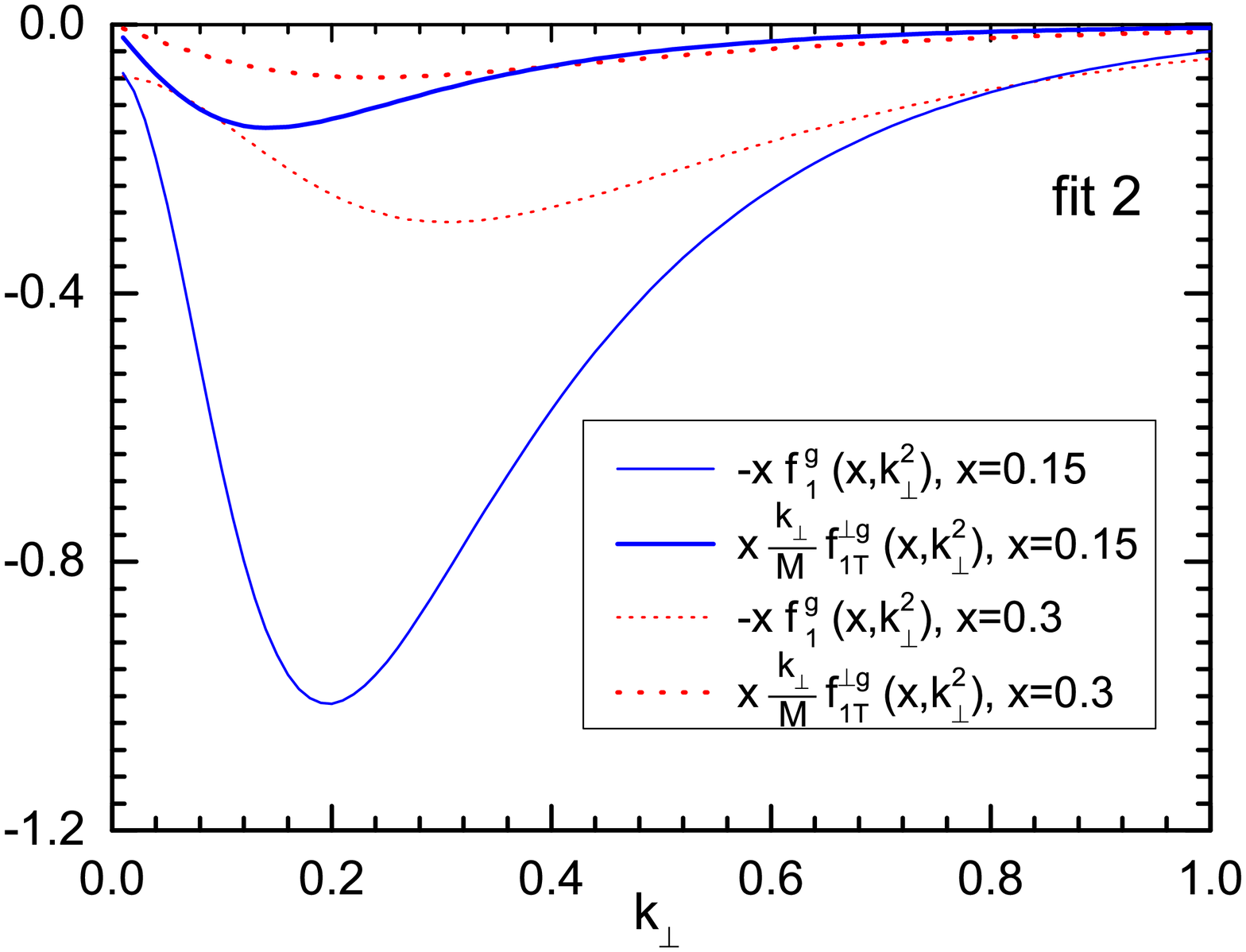}}
\caption{The $k_\perp$-dependence of the gluon TMD distribution functions in the spectator model at the model scale.
The thin and thick lines denotes $-xf_1^g(x,\bm k_\perp^2)$ and $x {k_\perp\over M}f_{1T}^{\perp g}(x,\bm k_\perp^2)$, while the solid and dotted lines show the results at $x=0.15$ and $0.3$, respectively.}
\label{fig:sivkt}
\end{figure*}

\begin{figure*}
\centering
\scalebox{0.35}{\includegraphics*[15pt,25pt][710pt,520pt]{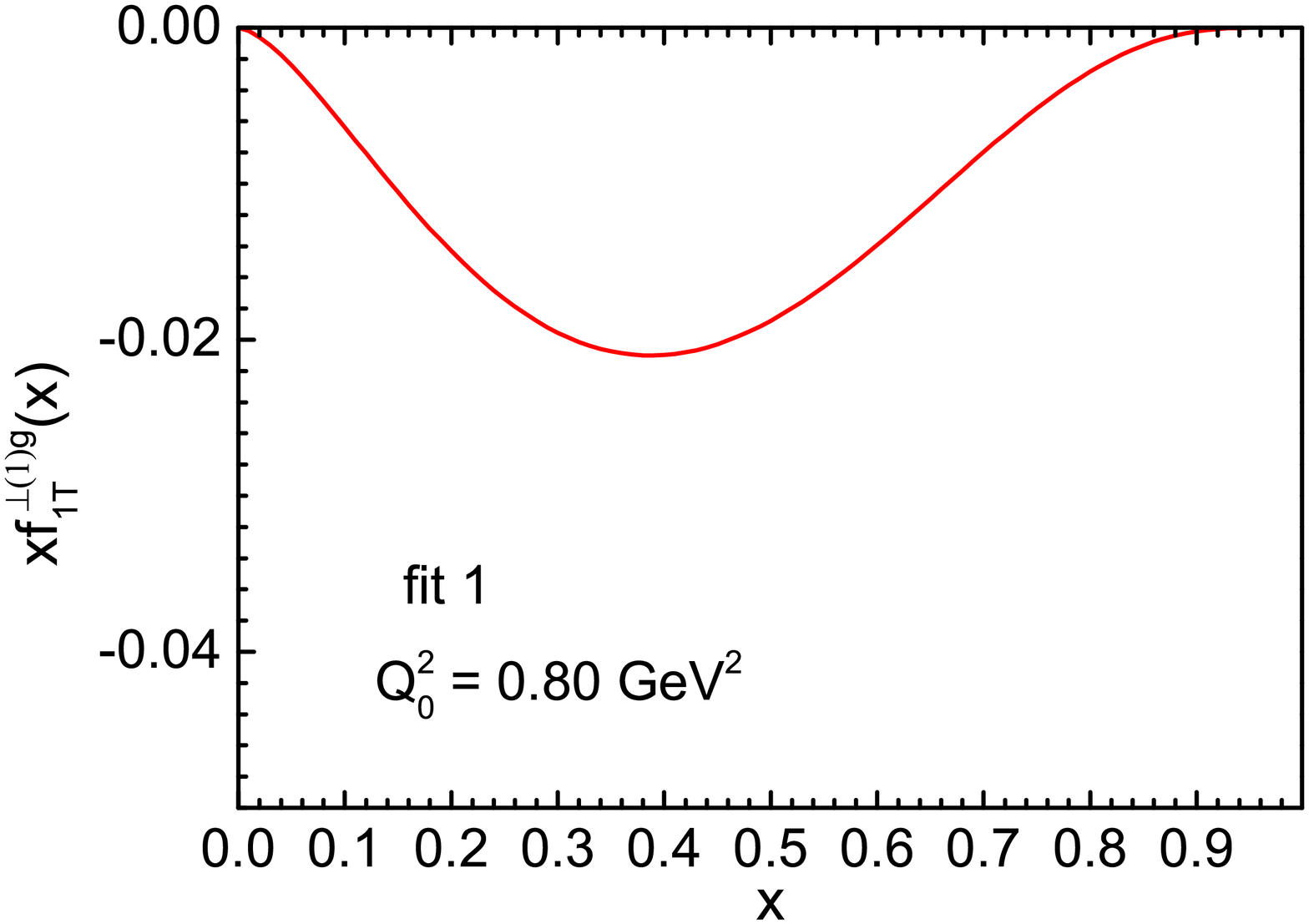}}~~
\scalebox{0.35}{\includegraphics*[15pt,25pt][710pt,520pt]{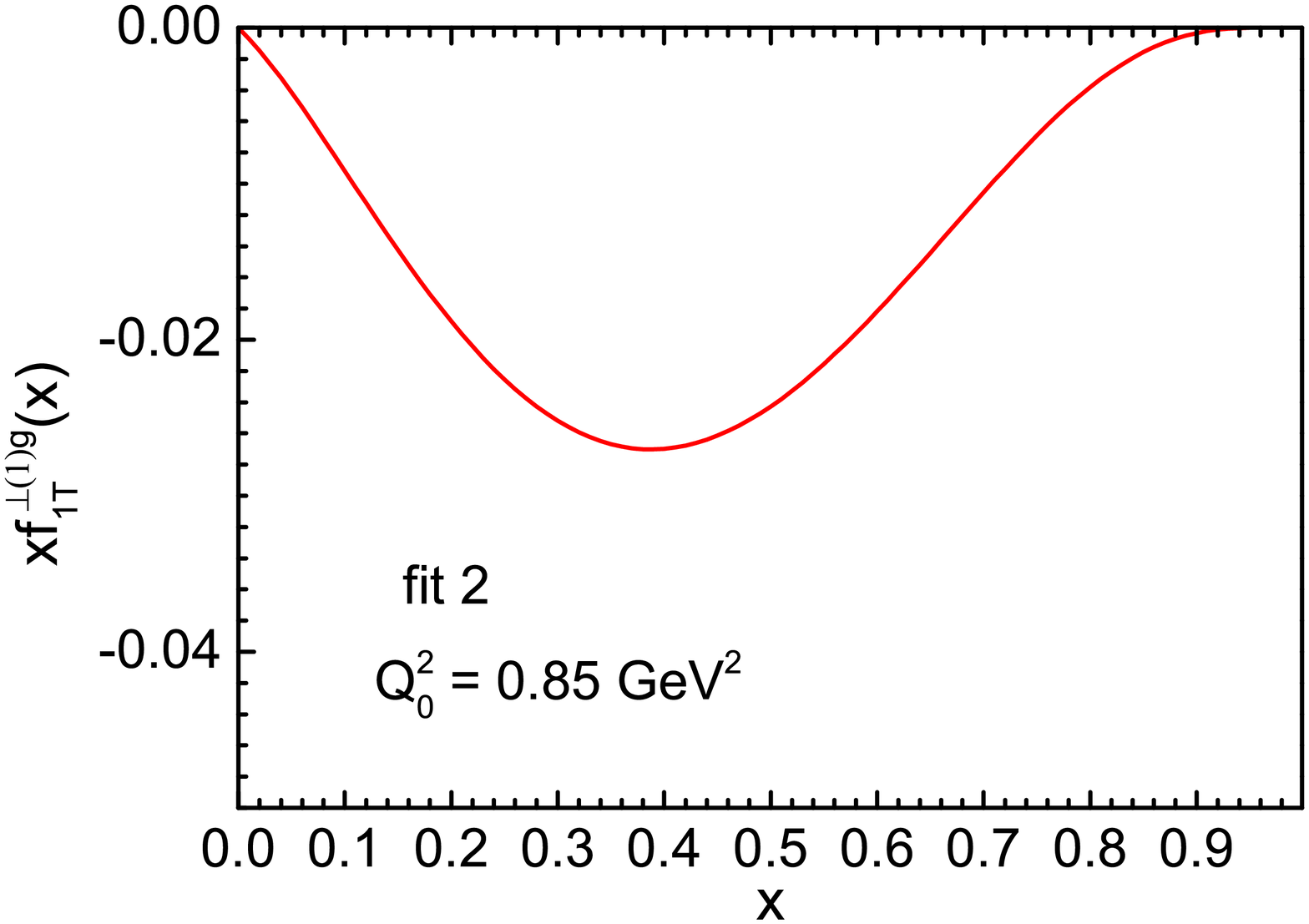}}
\caption{The first transverse moment of the gluon Sivers function $f_{1T}^{\perp (1)g}$ (timed with $x$) in the spectator model at the model scale.
The left and right panels correspond to the results calculated from the parameters obtained in fit 1 and fit 2, respectively.}
\label{fig:fitsiv}
\end{figure*}

Furthermore, when doing the fit, we have to choose an energy scale $Q_0^2$ at
which our model can be compared to the parametrization.
Here we consider the scale $Q_0^2$ as a special parameter of the model, which means that we also search the $Q^2$ region of the GRV98 PDF to find the lowest $\chi^2$.
The best results for the parameters from the two fits are shown in Table.~\ref{table1}.
In both fits we fix the parameter $M_g$ to be 0 GeV, since the gluon should be massless.
For the mass of the spectator, we required $M_X > M$, which is necessary to form a stable proton state.
In fit 1 we find that the lower possible scale is always preferred by the fit.
Therefore, in this fit we choose the lowest allowed scale of GRV98 parametrization as the model scale, which is $0.8 \textrm{GeV}^2$.
In fit 2, the model scale is slightly larger than the lowest allowed scale.

In Fig.~\ref{fig:f1g}, we plot the fitted gluon PDF $f_1^g(x)$ (solid line) in our model and compare it with the GRV98 parametrization (dashed line).
The left and right panels show the results from fit 1 and fit 2, respectively.
As the GRV98 parametrization does not provide the uncertainties for PDFs, we can not deduce from the fit the corresponding errors of the model parameters.
Here we assume that the error for our model resulting $f_1^g(x)$ is 30\% and show the error band in Fig.~\ref{fig:f1g}.
Although our spectator model is simple, we find that in both fits, the GRV98 gluon PDF can be well described by our model with 4 parameters.
One can also see that a better agreement between the model results and the parametrization is obtained in fit 1.

\begin{figure}
\centering
\scalebox{0.35}{\includegraphics{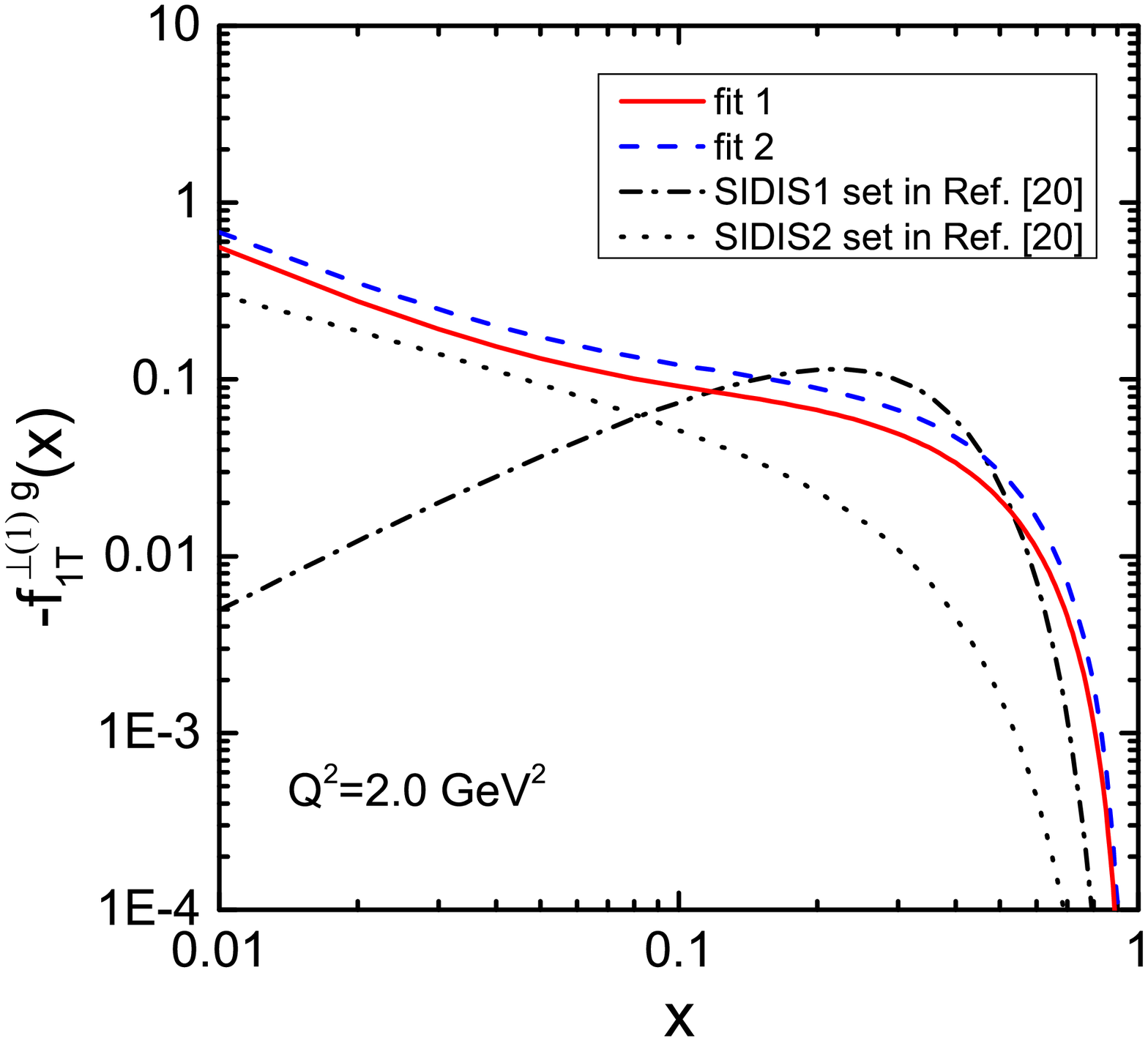}}~~
\caption{First transverse moment of the Sivers function compared with the extraction in Ref.~\cite{D'Alesio:2015uta}.}
\label{fig:sivq2}
\end{figure}

Using the values of the parameters obtained in the fits, we calculate the numerical result of the Sivers function at the model scale, and show $x{k_\perp\over M} f_{1T}^{\perp g}(x,\bm k_\perp^2)$ versus $k_\perp \equiv |\bm k_\perp|$ at $x=0.15$ and $0.3$ in Fig.~\ref{fig:sivkt}.
For comparison, we also show the $k_\perp$-dependence of $f_1^g(x,\bm k_\perp^2)$ (timed with $-x$).  
For the strong coupling $\alpha_S$ needed in the calculation, we adopt its value provided by the GRV98 code for consistency.
That is, we choose $\alpha_S^{\textrm{LO}}(Q_{0,\textrm{LO}}^{2}) = 0.47$ in fit 1 and $\alpha_S^{\textrm{NLO}}(Q_{0,\textrm{NLO}}^2)=0.41$ in fit 2, respectively.
We find that the $k_\perp$-dependence of the gluon TMD distribution functions changes when $x$ changes.
More specifically, the peak of the curves shifts from lower $k_\perp$ region to higer $k_\perp$ when $x$ increases.

The first transverse moment of the gluon Sivers function is defined as
\begin{align}
f_{1T}^{\perp (1)\, g}(x) = \int d^2\bm k_\perp \,{\bm k_\perp^2 \over 2M^2}\, f_{1T}^{\perp\, g} (x,\bm k_\perp^2) = -\Delta^N f_{g/p^\uparrow}^{(1)}(x).
\end{align}
Here the notion $\Delta^N f_{g/p^\uparrow}^{(1)}(x)$ is the one used in Ref.~\cite{D'Alesio:2015uta}.
In the left and right panels of Fig.~\ref{fig:fitsiv} we plot $xf_{1T}^{\perp (1)\, g}(x)$ as a function calculated form the parameters in fit 1 and fit 2, respectively.
We find that the first transverse moment of the gluon Sivers function in our model is negative, which is consistent with the result extracted in Ref.~\cite{D'Alesio:2015uta},
and $f_{1T}^{\perp (1)\, g}(x)$ calculated from fit 2 at the model scale is around 30\% larger than that from fit 1.

We also compare our model with the extracted gluon Sivers function in Ref.~\cite{D'Alesio:2015uta}.
Since the result in Ref.~\cite{D'Alesio:2015uta} is given at the scale $Q^2=2 \,\textrm{GeV}^2$, which is higher than our model scale, it is necessary to evolve the gluon Sivers function given at different scales to that at the same scale for comparison.
At the tree level, the $f_{1T}^{\perp(1)g}(x)$ can be related to the twist-3 tri-gluon function $T_G(x,x)$
\begin{align}
f_{1T}^{\perp(1)g}(x)\propto T_G(x,x)/M,
\end{align}
whereas the complete QCD evolution for $T_{G}(x,x)$ is given in Ref.~\cite{Braun:2009mi}.
However, its evolution is rather complicated, i.e., it also mix with the more general function $T_G(x,x^\prime)$ with $x\neq x^\prime $, the quark-gluon Qiu-Sterman function~\cite{qs} $T_F(x,x^\prime)$ and so on.
Here we only consider the homogenous term in the evolution to assume the following evolution kernel:
\begin{align}
P_{gg}(z) -N_c \delta(1-z)-N_c(1-z)\left(1+{1\over z}\right), \label{eq:kernel2}
\end{align}
where $P_{gg}(z)$ is the LO evolution kernel of the gluon to gluon splitting function for $f_1^g(x)$.
We expect that the approximation adopted here will not change the result qualitatively.
In Fig.~\ref{fig:sivq2}, we show the evolved $-f_{1T}^{\perp(1)g}(x)$ at $Q^2=2\,\textrm{GeV}^2$ from fit 1 and fit 2, and compare it with the gluon Sivers function extracted in Ref.~\cite{D'Alesio:2015uta} at the same scale.
We find that the evolution effect for $f_{1T}^{\perp(1)g}(x)$ is substantial.
In particular, the evolution from lower scale to higher scale increases the magnitude of $f_{1T}^{\perp (1)g}(x,Q^2)$ in the region $x<0.2$, while it decreases the magnitude of $f_{1T}^{\perp (1)g}(x,Q^2)$ in the larger $x$ region.
However, the scale dependence of $f_{1T}^{\perp(1)g}(x,Q^2)$ is weaker than that of $f_1^g(x,Q^2)$, because of the additional terms in Eq.~(\ref{eq:kernel2}).
We also find that the in the region $x<0.1$, the gluon sivers function from fit 1 is comparable with the SIDIS2 set in in Ref.~\cite{D'Alesio:2015uta} after the 10\% tolerance band is considered.
In the region $0.1<x<0.5$, the gluon sivers function from both fits qualitatively agrees with the SIDIS1 set in Ref.~\cite{D'Alesio:2015uta}.

Finally, we calculate the average transverse momentum of the gluon inside a transversely polarized proton:
\begin{align}
\langle  \bm k_\perp^g \rangle & = \int d^2 \bm k_\perp \bm k_\perp \Phi^g(x,\bm k_\perp;\bm S) \nonumber\\
&=-M\int_0^1 dx\, f_{1T}^{\perp(1)g}(x)\, ( \bm S \times \hat{\bm P})\nonumber\\
 &= \langle k_\perp^g \rangle\, ( \bm S \times \hat{\bm P}).
\end{align}
The quantity $\langle k_\perp^g \rangle$ can be constrained by the Burkardt sum rule and the average transverse momentum of quarks and antiquarks.
In Ref.~\cite{Anselmino:2008sga}, using the extraction of the Sivers distribution functions
for quarks and antiquarks, the authors provided a determination on the allowed range of $\langle k_\perp^g \rangle$ at $Q^2=2.4$ GeV$^2$:
\begin{align}
-10 \,\textrm{MeV}<\langle  k_\perp^g \rangle <48\, \textrm{MeV}.
\end{align}
We evolve the gluon Sivers function in our model to the scale $Q^2=2.4$ GeV$^2$ and estimate $\langle k_\perp^g \rangle$ in the two fits
\begin{align}
\textrm{fit 1}:~\,\langle k_\perp^g \rangle = 38 \,\textrm{MeV};~~~\textrm{fit 2}:~\,\langle k_\perp^g \rangle = 51\,\textrm{MeV}. \nonumber
\end{align}
Our results show that the average transverse momentum of the gluon from fit 1 agrees with the bound on $\langle k_\perp^g \rangle$ given in Ref.~\cite{Anselmino:2008sga}.

A very important theoretical constraint on the Sivers function is the positivity bound~\cite{Bacchetta:1999kz,Mulders:2000sh}
\begin{align}
{k_\perp \over M}\,\left|f_{1T}^{\perp\, g}(x,\bm k_\perp^2)\right| \le f_1^g(x,\bm k_T^2). \
\end{align}
We have checked that the gluon Sivers function from our model satisfies the above inequality in the region $k_\perp \equiv |\bm k_\perp| < 2$ GeV.
This is similar to the quark sector of the Sivers function in the quark-diquark model, for which a violation of the positivity bound is observed~\cite{Kotzinian:2008fe} at high $k_\perp$.
As explained in Ref~\cite{Pasquini:2011tk}, the violation of the inequality for T-odd TMDs may be due to the fact that T-odd TMDs is evaluated to $\mathcal{O}(\alpha_s)$, while T-even TMD distributions is truncated at $\mathcal{O}(\alpha_s^0)$ in model calculations.
However, we find that in our model the gluon TMD distributions are very small (less than $10^{-4}$) at the region the positivity bound is violated.
Besides, $2$ GeV is much larger than the mean transverse momentum of the gluon in our model.
Therefore, the fact that in our model the inequality only holds in the region $k_\perp < 2 $ GeV is an acceptable result, as our model is assumed to be valid in the region where $k_\perp$ is not so large.

\section{Conclusion}

In this work, we studied the gluon Sivers function using a light-cone spectator model.
We treat the Fock state of the proton state as a composite system formed by a gluon and a spin-$1/2$ spectator particle, in the case the active parton is a gluon.
Using the overlap representation, we calculated the unpolarized gluon distribution function $f_1^g(x,\bm k_\perp^2)$ and the gluons Sivers function $f_{1T}^{\perp}(x,\bm k_\perp^2)$.
In the calculation, we adopte the wavefunctions of the proton Fock state motivated by the wavefunctions of the electron Fock state.
Besides, we choose the Brodsky-Huang-Lepage prescription for the wavefunctions in the momentum space.
Furthermore, in the case of the Sivers function, we adopt an interaction kernel that simulates the final-state interaction between the gluon and the spectator.
The values of the parameters in the model are determined by fitting the model resulting $f_1^g(x)$ with the GRV98 parametrization.
Specifically, the LO and NLO gluon PDF sets are adopted to obtain two sets of fit.
Our numerical calculations show that the first transverse moment of the gluon Sivers function $f_{1T}^{\perp(1)g}$ in our model is negative, and is several percent in magnitude.
We also compared our model results with the recent extraction of $f_{1T}^{\perp(1)g}$ and find that our result can coincide with the parametrization of $f_{1T}^{\perp(1)g}$ in Ref.~\cite{D'Alesio:2015uta}.
In the estimate we taken into account the evolution of the gluon Sivers function, and
find that it is important to include the evolution effect in order to compare model calculation with phenomenological analysis given at different energy scales.
Finally, our model can be suitably extended to estimate the gluon helicity distribution and we leave it as a future study.
In conclusion, our study may provide useful information of the gluon Sivers function from an intuitive model concerning the gluon structure of the nucleon.

\section*{Acknowledgements}
This work is partially supported by the National Natural Science
Foundation of China (Grants No.~11575043, No.~11120101004 and No.~11475006), and
by the Qing Lan Project.

\end{document}